\documentclass{article}
\usepackage{spconf,amsmath,graphicx}
\usepackage{tabularx}
\usepackage[english]{babel}
\usepackage{multirow}
\usepackage{multicol}
\usepackage{enumitem}
\usepackage{hyperref}
\usepackage{diagbox}
\usepackage[belowskip=-10pt,aboveskip=5pt]{caption}

\setlist[enumerate]{itemsep=-1mm}

\title{Deep Mouse: An End-to-end Auto-context Refinement Framework for Brain Ventricle \& Body Segmentation in Embryonic Mice Ultrasound Volumes}
\name{\begin{tabular}{c} Tongda Xu$^{1}$\textsuperscript{*}, Ziming Qiu$^{1}$\textsuperscript{*}, William Das$^{2}$, Chuiyu Wang$^{3}$, Jack Langerman$^{4}$, Nitin Nair$^{1}$, Orlando\\ Aristiz\'{a}bal$^{5,6}$, Jonathan Mamou$^{5}$, Daniel H. Turnbull$^{6}$, Jeffrey A. Ketterling$^{5}$, Yao Wang$^{1}$ \thanks{The research described in this paper was supported in part by NIH grant EB022950 and HD097485. \textsuperscript{*} Authors contributed equally.} \end{tabular}}
\address{$^{1}$Department of Electrical and Computer Engineering, New York University, New York, USA\\
$^{2}$Hunter College High School, New York, USA\\
$^{3}$School of Electronic and Information Engineering, Beihang University, Beijing, China\\
$^{4}$Department of Computer Science, New York University, New York, USA\\
$^{5}$F. L. Lizzi Center for Biomedical Engineering, Riverside Research, New York, USA\\
$^{6}$Skirball Institute of Biomolecular Medicine, New York University, New York, USA
}
\begin{document}
%

\maketitle
\begin{abstract}


High-frequency ultrasound (HFU) is well suited for imaging embryonic mice due to its noninvasive and real-time characteristics. However, manual segmentation of the brain ventricles (BVs) and body requires substantial time and expertise. This work proposes a novel deep learning based end-to-end auto-context refinement framework, consisting of two stages. The first stage produces a low resolution segmentation of the BV and body simultaneously. The resulting probability map for each object (BV or body) is then used to crop a region of interest (ROI) around the target object in both the original image and the probability map to provide context to the refinement segmentation network. Joint training of the two stages provides significant improvement in Dice Similarity Coefficient (DSC) over using only the first stage (0.818 to 0.906 for the BV, and 0.919 to 0.934 for the body). The proposed method significantly reduces the inference time (102.36 to 0.09 s/volume $ \approx $1000x faster) while slightly improves the segmentation accuracy over the previous methods using slide-window approaches. 


\end{abstract}
\section{Introduction}
\label{sec:intro}

The mouse, due to its high degree of homology with human genome, is widely used for studies of embryonic mutations. The physical expression of genetic mutations can be identified in terms of variations in the shape of the brain ventricle (BV) or other parts of the body \cite{kuo2015automatic}. High-frequency ultrasound (HFU) is well suited for imaging embryonic mice because it is noninvasive, real-time and can still provide fine-resolution volumetric images \cite{aristizabal2013high}. Manual segmentation of the BV and body is time-consuming necessitating the development of fully automatic and real-time segmentation algorithms \cite{henkelman2010systems}.

Earlier works attempted to address segmentation of volumetric embryonic data. Nested Graph Cut (NGC) \cite{kuo2015nested} was developed to perform the segmentation of the BV from a HFU mouse embryo head image manually cropped from a whole-body scan, and \cite{kuo2018automatic} extended to perform BV segmentation in whole-body images, which worked well on a small data set but failed to generalize to a larger unseen data set. 
\begin{figure}[t!]
    \centering
    \includegraphics[width=\linewidth]{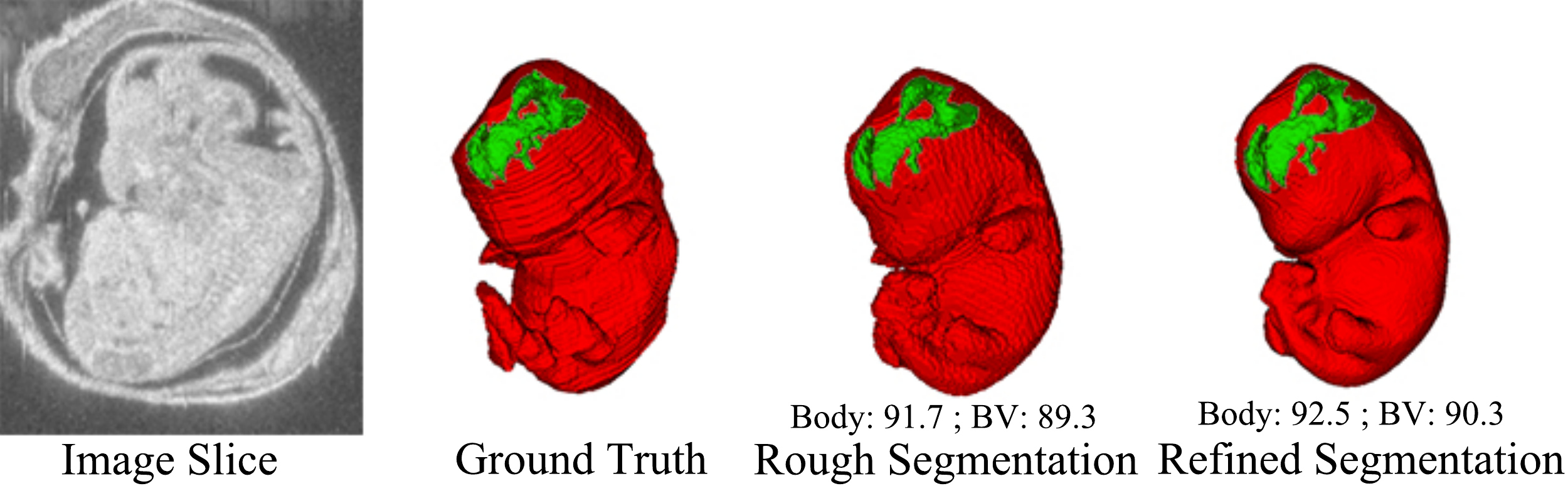}
    \caption{An example of image data, ground truth label, initial rough segmentation and final refined segmentation. The numbers below the 3D segmentation are corresponding DSC.}
    \label{fig:intro}
\end{figure}

\begin{figure*}[th]
  \includegraphics[width=\textwidth]{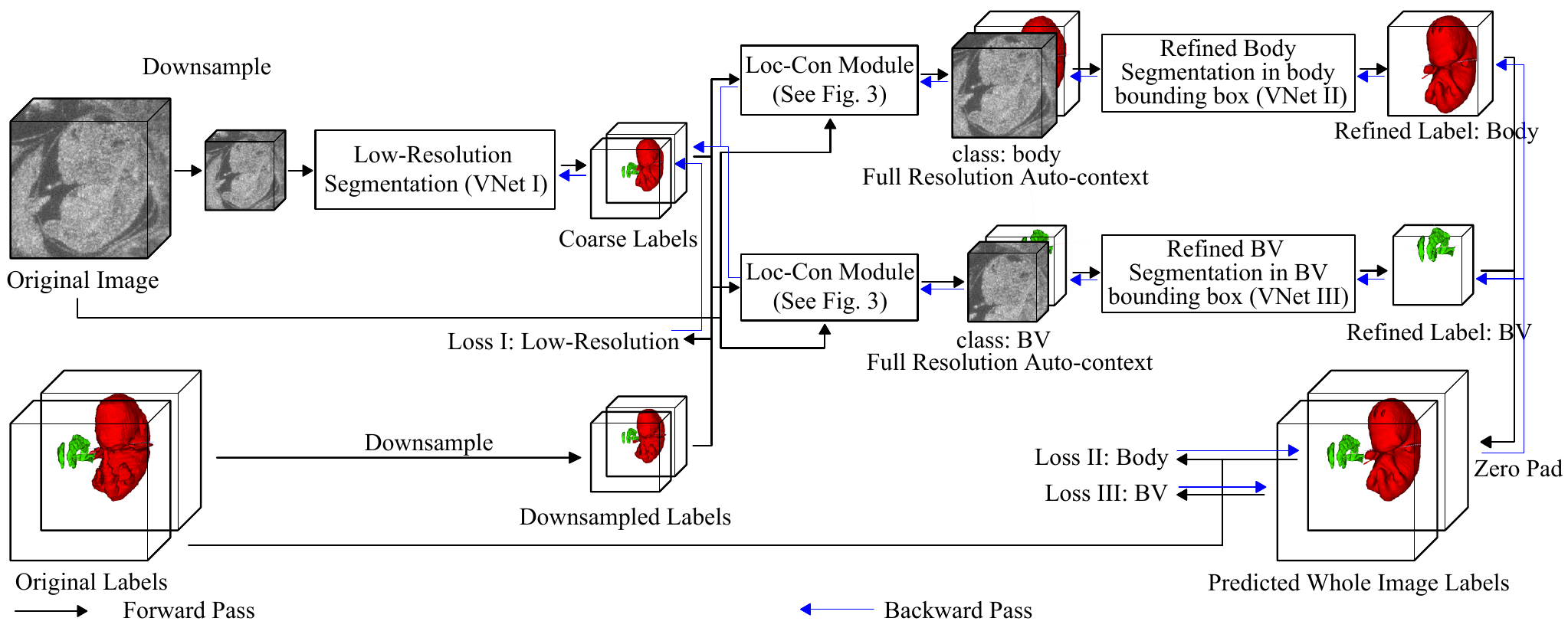}
  \caption{Diagram of overall pipeline.}
  \label{fig:pipline}
\end{figure*}

\begin{figure}[th]
    \centering
    \includegraphics[width=\linewidth]{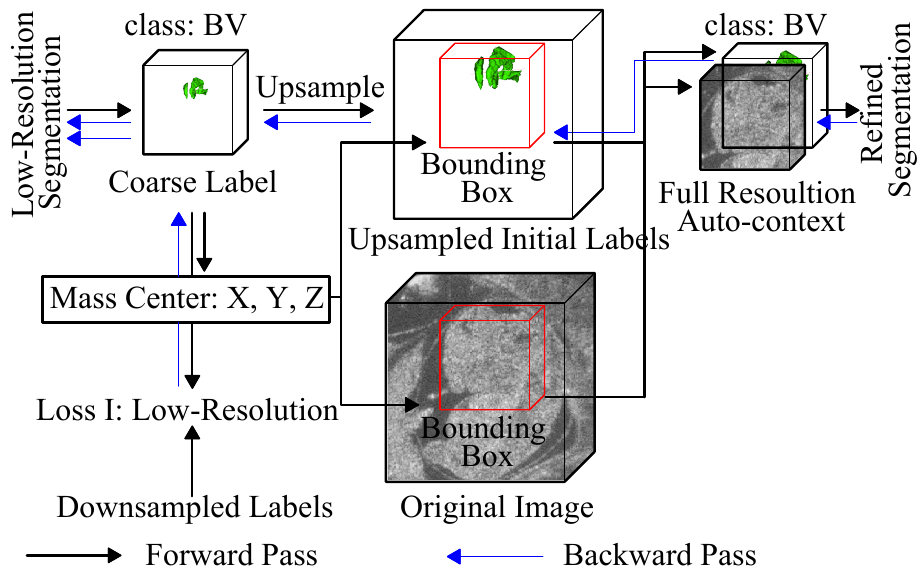}
    \caption{Diagram of localization-auto-context (Loc-Con) module for the BV. A similar configuration is used for the body. The gradient produced by the refinement loss can flow back to low resolution segmentation network (blue arrows).}
    \label{fig:locat}
\end{figure}

Inspired by the success of fully convolutional networks (FCN) for semantic segmentation \cite{long2015fully}, a deep-learning based framework for BV segmentation was proposed in \cite{qiu2018deep} which outperformed the NGC based framework in  \cite{kuo2018automatic} by a large margin. Because the BV makes up a very small portion ($<$0.5\%) of the whole volume, the algorithm \cite{qiu2018deep} first applied a volumetric convolutional neural network (CNN) on a 3D sliding window over the entire volume to identify a 3D bounding box containing the entire BV, followed by a FCN to segment the detected bounding box into BV or background. However, despite high accuracy (0.904 DSC for BV), hundreds of thousands of forward passes of a classification network is required. The challenges for body segmentation are similar to those for BV segmentation, except the extreme imbalance between foreground and background is somewhat alleviated (the body makes up around 10\% of the whole volume). Hence, the localization step is not necessary for body segmentation. Qiu et al. \cite{qiu2019automatic} used the same FCN for BV segmentation in \cite{qiu2018deep} to segment the body in a sliding window based manner. This sliding window based method is inefficient for the same reason as the localization network in \cite{qiu2018deep}. The BV is contained inside the head of the body, which makes it possible to segment both of them simultaneously in a unified and efficient framework.

Roth et al. \cite{roth2018application} focused on abdominal CT image segmentation. They applied two cascaded 3D FCNs using the initial segmentation results to localize the foreground organs and reduced the size of the 3D region which was input to the second FCN. The initial segmentation was used only for localization and was not concatenated with the raw image as input to the second FCN. Tang et al. \cite{tang2019multi} cascaded four UNets and trained them in an end-to-end manner for skin lesion segmentation. However, this framework did not use the segmentation output of previous UNet to reduce the spatial input to the next UNet and was restricted to 2D binary segmentation.

Here, we propose an efficient end-to-end auto-context refinement framework for joint BV and body segmentation from volumetric HFU images (Fig.~\ref{fig:pipline}). The idea behind auto-context \cite{tu2008auto} is to iteratively approach the ground truth by a sequence of models where the input and output of the previous model is concatenated to form the input for the next model such that the final segmentation is closer to the ground truth than the intermediate segmentation (Fig.~\ref{fig:intro}). Specifically, a VNet like network (VNet I) \cite{milletari2016v} was first applied to a down-sampled HFU 3D image to jointly segment BV and body. The resulting low resolution BV label was then up-sampled to the original resolution, and a bounding box containing the BV was generated (Fig. \ref{fig:locat}). Next, the original image and the initial BV predicted probability map in the bounding box were concatenated as localized auto-context input and fed into another VNet (VNet III) to generate the refined BV label. A parallel process was used to generate final fine-resolution body segmentation using a third VNet (VNet II). Each VNet was trained separately and then fine-tuned in an end-to-end manner. Compared with other works, this framework has the following advantages:

\begin{enumerate}
    \item The class imbalance problem posed by segmenting small structures in large volumes and the memory issue associated with large volumetric images were mitigated by cascading the networks from low resolution for the whole image to high resolution in the localized region.
    
    \item The auto-context provided by concatenating the up-sampled low resolution predicted probability maps with the full resolution images allows efficient full resolution segmentation which can incorporate information from a wide field of view.
    
    \item The combination of the two stages in a pipeline allows end-to-end training and efficient, real-time one-pass inference while achieving segmentation accuracy slightly better than the sliding-window based approach, which is the substantially more time consuming.
    
\end{enumerate}

\section{Methods}
\label{sec:meth}

The overview of our proposed end-to-end BV and body segmentation framework is shown in Fig.~\ref{fig:pipline}. The pipeline consists of two stages: (1) initial segmentation and (2) segmentation refinement. The initial segmentation produces a joint low resolution segmentation maps for BV and body simultaneously. Next, the original data and the low resolution label for each object are passed to a localization-auto-context (Loc-Con) module (Fig.~\ref{fig:locat}), which generates a bounding box for the object based on the centroid of the up-sampled predicted probability map, and concatenates the original (full-resolution) image and initial up-sampled predicted probability map as auto-context input for the refinement network for this object. Then, the refinement network generates a refined full resolution segmentation map within the bounding box. Finally, all the independently trained networks are jointly fine-tuned to further improve the segmentation performance.

\subsection{Initial segmentation on low resolution}
\label{sec:sec:meth_1}

Because of memory constraints, the low resolution images were rescaled and processed at a size of $160^3$ voxels. A VNet-like \cite{milletari2016v} structure (VNet I) was trained to perform BV and body segmentation simultaneously at low resolution. The output of the VNet had 3 channels, representing the background, BV, and body. The Dice loss \cite{milletari2016v} for each class was summed and used as the training loss (loss I in Fig. \ref{fig:pipline}).

\subsection{Localization-auto-context module}
\label{sec:sec:meth_2}

To better utilize the information given from the low resolution segmentation result for each object, the Loc-Con module was introduced to generate the localized auto-context input for refinement. For each foreground object (BV or body), Loc-Con module steps were (Fig.~\ref{fig:locat}): 
\begin{enumerate}
    \item Up-sample (trilinear interpolation) the initial segmentation map to original resolution.
    \item Generate a fixed size bounding box ($160^3$ for BV and $256^3$ for body) located at the center of predicted probability map for each class. 
    \item Concatenate the original resolution image and initial predicted probability map in the bounding box to create the auto-context input for the refinement network.
\end{enumerate}

Going beyond the previous works \cite{roth2018application}\cite{tang2019multi}, the Loc-Con module acts as an attention mechanism by leveraging the low resolution rough segmentation to crop a ROI (the bounding box) at the original resolution. It also draws on the conventional auto-context strategy \cite{tu2008auto} by providing an initial predicted probability map as a separate input channel. This initial map obtained from the whole image at the low resolution provides global context information, which helps to improve the final segmentation results. It also enables the gradient from the refinement networks (VNet II $\&$ III) to flow back to the initial segmentation network (VNet I), which makes end-to-end fine-tuning feasible (Fig. \ref{fig:pipline}).

\subsection{Pre-training of fine resolution refinement network}

Two refinement networks (VNet II $\&$ III) were trained for BV and body, respectively. For each object, the high resolution raw image and the up-sampled initial segmentation probability map in the localized bounding box were concatenated and used as the input. The structure of the refinement network was exactly the same as the initial segmentation, except that it takes 2 channels as input and produces 1 channel as output. Using the object centroid information, the output was zero-padded back to original image size.

\subsection{End-to-end refinement on fine resolution}
\label{sec:sec:meth_3}

During the pretraining of the refinement stage, the parameters of the initial segmentation network (VNet I) were frozen until the refinement network for each object (VNet II $\&$ III) converged. After that, all three networks were jointly optimized to minimize the sum of Dice losses measured on the fine resolution image (loss II $\&$ III in Fig. \ref{fig:pipline}). 

\section{EXPERIMENTS}
\label{sec:experi}
\subsection{Data description and implementation details}
\label{sec:experi_1}

The data set used in this work consisted of 231 HFU mouse embryo volumes which were acquired in utero and in vivo from pregnant mice (10-14.5 days after mating) using a 5-element 40-MHz annular array \cite{aristizabal2013high}. The dimensions of the HFU volumes vary from 150 $\times$ 161 $\times$ 81 to 210 $\times$ 281 $\times$ 282 voxels and the voxel size is 50 $\times$ 50 $\times$ 50 $\mu$m. For each of the 231 volumes, manual BV and body segmentations were conducted by trained research assistants using Amira, a commercial software. The data were then randomly split into 185 for training and 46 for testing.

All the neural network models were implemented in PyTorch 1.2 \cite{paszke2017automatic}, with CUDA 9.1 using two NVIDIA Tesla P40s. To compensate for limited data, original images were randomly rotated from $-180^{\circ}$ to $180^{\circ}$ along each of the three axes, then randomly translated $-30$ to $30$ voxels and finally randomly flipped. During the initial pretraining step (VNet I), the Dice loss (loss I) between the predicted segmentations and the ground truth labels was averaged across the three classes (background, body and BV). During the pre-train refinement stage (VNet II $\&$ III) and end-to-end refinement stage (VNet I, II $\&$ III), the Dice loss for body and BV (loss II $\&$ III) were used to train the networks. All networks were trained with the Adam optimizer \cite{kingma2014adam} with learning rate $10^{-2}$ for the initial segmentation and pre-train refinement stages. A learning rate of $10^{-3}$ was used for the end-to-end refinement stage.

\subsection{Results and discussion}
\label{sec:experi_2}

\begin{table}[bh]
\caption{The Dice Similarity Coefficient (DSC) and inference time averaged over 46 test volumes for different methods.} \label{tab:tabel_result}
\begin{tabularx}{\linewidth}{|>{\hsize=1.4\hsize\linewidth=\hsize}X|l|l|>{\hsize=.6\hsize\linewidth=\hsize}X|} \hline
\backslashbox{Methods}{Results} &BV DSC &Body DSC   & Inference \newline Time $^*$ \\ \hline
                                 Benchmark                   & 0.904 \cite{qiu2018deep}       & 0.932 \cite{qiu2019automatic} & 102.36s    \\ \hline
Initial  Segmentation      & 0.818  & 0.918  & 0.006s    \\ \hline
Refinement w/o \newline Auto-Context Input & 0.878 & 0.922     & 0.08s    \\ \hline
Refinement  w/ \newline Auto-Context Input & 0.894 & 0.924 & 0.09s \\ \hline
\textbf{Refinement \newline End-to-end}    & \textbf{0.906}  & \textbf{0.934}       &  0.09s \\ \hline
\end{tabularx}
     {\raggedright \small{* The average inference time is calculated on two NVIDIA Tesla P40 graphic cards. The benchmark inference time is summed over separate BV \cite{qiu2018deep} and body \cite{qiu2019automatic} segmentations.} \par}
\end{table}

\begin{figure}[t!]
    \centering
    \includegraphics[width=\linewidth]{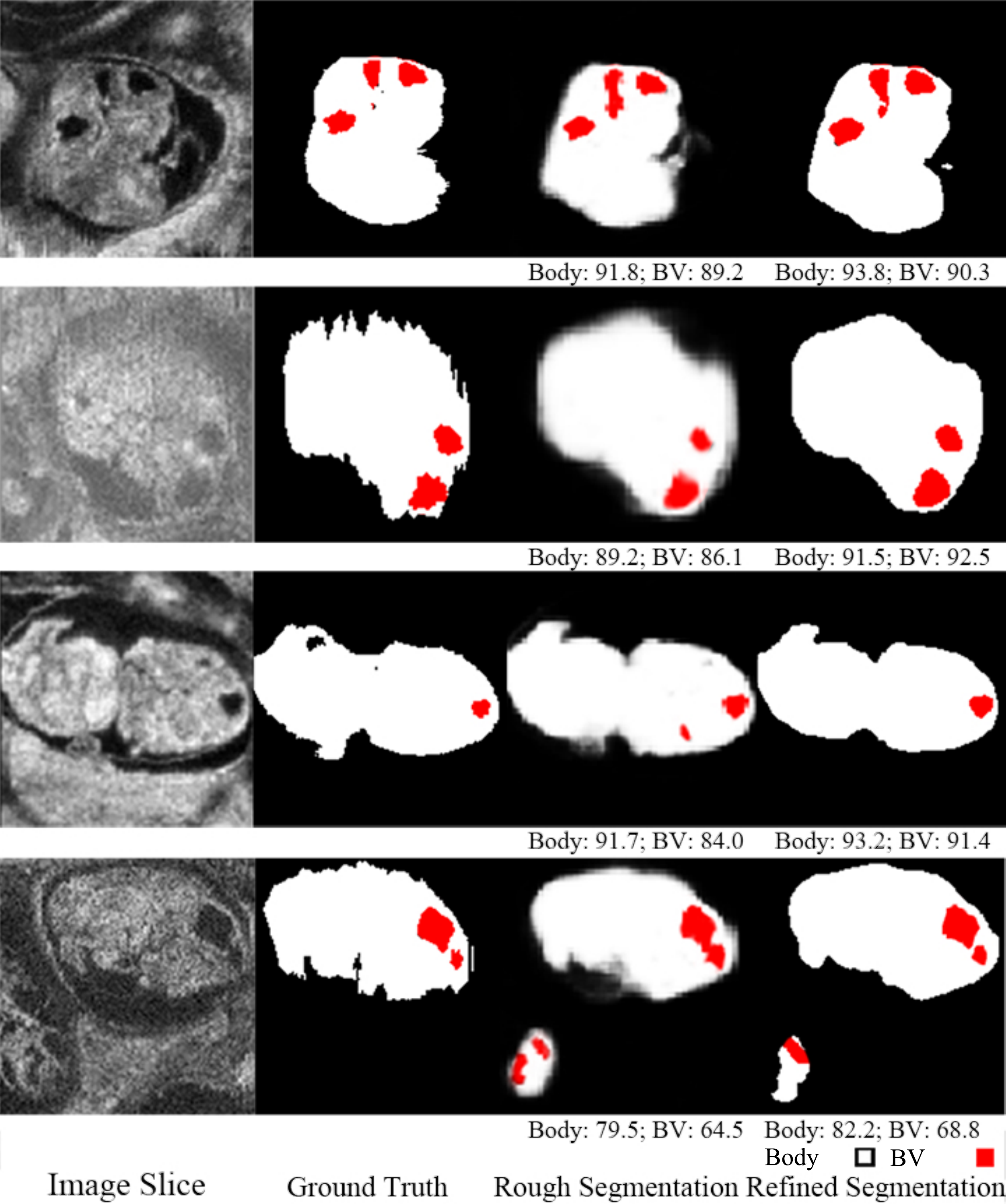}
    \caption{Qualitative segmentation results of end-to-end auto-context refinement framework for 4 ultrasound volumes. Red indicates BV, white indicates body and the numbers below the predicted segmentation are corresponding DSC. The second row is an image with motion artifacts so the ground truth is noisy in the body background boundary. The refined network produces a smooth boundary which is closer to the true physical structure. The last row is an image with one complete embryo and one partial embryo but only the complete one is labeled. The predicted label contained part of the incomplete embryo.}
    \label{fig:sample}
\end{figure}

As shown in Table.~\ref{tab:tabel_result}, the initial segmentation achieves average DSC of 0.818 and 0.918 for BV and body respectively. The results on the body are still competitive, while the BV segmentation performance is unsatisfactory. This phenomenon is as expected due to the fact that BV is much smaller than the body. Hence, it is necessary to localize the ROI and refine the segmentation.

In order to determine the effectiveness of the auto-context approach, refinement without concatenation only inputs the raw image cropped from the bounding box found in the localization step to the refinement network (without including the initial predicted probability map as a second channel). Compared to using the initial segmentation and the raw image (auto-context), this approach suffers from a significant degradation in DSC for the BV (from 0.894 to 0.878 DSC). Finally, end-to-end refinement improves BV DSC to 0.906 and body to 0.934. Although the performance is only slightly better than the previous sliding-window based methods \cite{qiu2018deep}\cite{qiu2019automatic}, the new method has significantly shorter inference time (0.09 second for each image, compared to 102.36 second). For fair comparison, the networks from \cite{qiu2018deep} and \cite{qiu2019automatic} were retrained using the same training set described here and evaluated on the same testing set.

    As shown in Fig.~\ref{fig:sample}, the initial segmentation produces reasonable body segmentations along with very rough BV segmentations. After end-to-end refinement, the segmentation accuracy was substantially improved.

\section{CONCLUSION}
In this work, an end-to-end auto-context refinement framework was proposed consisting of two stages. The initial segmentation acts not only as a ROI localization module, but also provides global context information when fed as the second channel together with the original images to the network. Experiments demonstrate the necessity of this two-stage structure and the effectiveness of end-to-end fine-tuning. The proposed method achieves DSC of 0.906 and 0.934 for BV and body segmentation respectively, outperforming the previous methods in accuracy, while also being around a thousand times faster. In conclusion, the proposed algorithm could be invaluable for phenotyping studies.


\bibliographystyle{IEEEbib}
\bibliography{refs}

\end{document}